\documentclass[
  journal=largetwo,
  manuscript=article-type,
  year=2020,
  volume=37,
]{cup-journal}
\usepackage{svg}
\usepackage{amsmath}
\usepackage{amssymb}
\usepackage{booktabs}
\usepackage{subcaption}
\usepackage{physunits}
\newcommand{\HI}{\mbox{\sc H{i}}}
\title{WALLABY Pilot Survey: HI source-finding with a machine learning framework}

\author{Li Wang}
\affiliation{ATNF, CSIRO Space and Astronomy, P.O. Box 1130, Bentley, WA 6102, Australia}
\email[Li Wang]{Li.Wang1@csiro.au}

\author{O. Ivy\ Wong}
\affiliation{ATNF, CSIRO Space and Astronomy, P.O. Box 1130, Bentley, WA 6102, Australia}
\alsoaffiliation{International Centre for Radio Astronomy Research (ICRAR), University of Western Australia, 35 Stirling Hwy, Crawley, WA 6009, Australia}
\alsoaffiliation{ARC Centre of Excellence for All-Sky Astrophysics in 3 Dimensions (ASTRO 3D), Australia}

\author{Tobias\ Westmeier}
\affiliation{International Centre for Radio Astronomy Research (ICRAR), University of Western Australia, 35 Stirling Hwy, Crawley, WA 6009, Australia}
\alsoaffiliation{ARC Centre of Excellence for All-Sky Astrophysics in 3 Dimensions (ASTRO 3D), Australia}

\author{Chandrashekar\ Murugeshan}
\affiliation{ATNF, CSIRO Space and Astronomy, P.O. Box 1130, Bentley, WA 6102, Australia}
\alsoaffiliation{ARC Centre of Excellence for All-Sky Astrophysics in 3 Dimensions (ASTRO 3D), Australia}

\author{Karen\ Lee-Waddell}
\affiliation{International Centre for Radio Astronomy Research (ICRAR), University of Western Australia, 35 Stirling Hwy, Crawley, WA 6009, Australia}
\alsoaffiliation{ATNF, CSIRO Space and Astronomy, P.O. Box 1130, Bentley, WA 6102, Australia}
\alsoaffiliation{International Centre for Radio Astronomy Research (ICRAR), Curtin University, Bentley, WA 6102, Australia}

\author{Yuanzhi.\ Cai}
\affiliation{CSIRO Mineral Resource, 26 Dick Perry Ave, Kensington, WA 6151, Australia}

\author{Xiu.\ Liu}
\affiliation{ATNF, CSIRO Space and Astronomy, P.O. Box 1130, Bentley, WA 6102, Australia}
\alsoaffiliation{Western Australian School of Mines: Minerals, Energy and Chemical Engineering, Curtin University, GPO Box U1987, Perth, WA 6845, Australia}

\author{Austin Xiaofan\ Shen}
\affiliation{ATNF, CSIRO Space and Astronomy, P.O. Box 1130, Bentley, WA 6102, Australia}

\author{Jonghwan\ Rhee}
\affiliation{International Centre for Radio Astronomy Research (ICRAR), University of Western Australia, 35 Stirling Hwy, Crawley, WA 6009, Australia}

\author{Helga\ Dénes}
\affiliation{School of Physical Sciences and Nanotechnology, Yachay Tech University, Hacienda San José S/N, 100119, Urcuquí, Ecuador}

\author{Nathan\ Deg}
\affiliation{Department of Physics, Engineering Physics, and Astronomy, Queen's University, Kingston ON K7L~3N6, Canada}

\author{Peter\ Kamphuis}
\affiliation{Ruhr University Bochum, Faculty of Physics and Astronomy, Astronomical Institute (AIRUB), 44780 Bochum, Germany}

\author{Barbara\ Catinella}
\affiliation{International Centre for Radio Astronomy Research (ICRAR), University of Western Australia, 35 Stirling Hwy, Crawley, WA 6009, Australia}
\alsoaffiliation{ARC Centre of Excellence for All-Sky Astrophysics in 3 Dimensions (ASTRO 3D), Australia}

\keywords{data analysis, radio lines, surveys} 

\newcommand{\red}[1]{\textcolor{black}{#1}}

\begin{document}

\begin{abstract}
The data volumes generated by the WALLABY atomic Hydrogen (\HI) survey using the Australian Square Kilometre Array Pathfinder (ASKAP) necessitate greater automation and reliable automation in the task of source-finding and cataloguing.  To this end, we introduce and explore a novel deep learning framework for detecting low Signal-to-Noise Ratio (SNR) \HI\ sources in an automated fashion.  Specifically, our proposed method provides an automated process for separating true \HI\ detections from false positives when used in combination with the Source Finding Application (SoFiA) output candidate catalogues.  Leveraging the spatial and depth capabilities of 3D Convolutional Neural Networks (CNNs), our method is specifically designed to recognize patterns and features in three-dimensional space, 
making it uniquely suited for rejecting false positive sources in low SNR scenarios generated by conventional linear methods. As a result, our approach is significantly more accurate in source detection and results in considerably fewer false detections compared to previous linear statistics-based source finding algorithms. Performance tests using mock galaxies injected into real ASKAP data cubes reveal our method's capability to achieve near-100\% completeness and reliability at a relatively low integrated SNR $\sim3-5$. 
An at-scale version of this tool will greatly maximise the science output from the upcoming widefield \HI\ surveys. 
\end{abstract}

\section{Introduction}
The Widefield ASKAP L-band Legacy All-sky Blind surveY \citep[WALLABY; ][]{koribalski2020wallaby} using the Australian SKA Pathfinder (ASKAP)  is expected to map a large portion of the southern sky in the 21-cm line emission of neutral hydrogen (\HI).  WALLABY expects to detect \HI\ from over 200,000 galaxies out to a redshift of $z \approx 0.1$, amounting to approximately 1 petabyte in data volume.

Given the very large amount of imaging data anticipated from  WALLABY, the detection and characterisation of galaxies will need to occur in a fully automated fashion with minimal manual intervention.  To this end, dedicated \HI\ source finding software such as DUCHAMP \citep{whiting2012duchamp}, SELAVY \citep{whiting2012source}, and the Source Finding Application \citep[SoFiA; ][]{serra2015SoFiA} have been developed.  SoFiA encapsulates the outcomes and technical knowledge from previous generations of large \HI\ surveys and their development of automated source-finding methods \citep{popping12}.  Parallel processing and multithreading has been built into SoFiA2 \citep{westmeier2021SoFiA} to enable more efficient source-finding in very large \HI\ survey datasets, such as those from WALLABY.

However, it is imperative to acknowledge the limitations of current automated methodologies, especially in the case of non-Gaussian noise characteristics. At low Signal-to-Noise Ratios (SNR $<5$), these algorithms are susceptible to generating significant numbers of false detections. The manual vetting required to separate false positives from true \HI\ detections, especially in the context of WALLABY's extensive dataset, poses a considerable challenge and bottleneck in efficiency. 

\begin{figure*}[ht]
\centering
\includegraphics[width=0.85\linewidth]{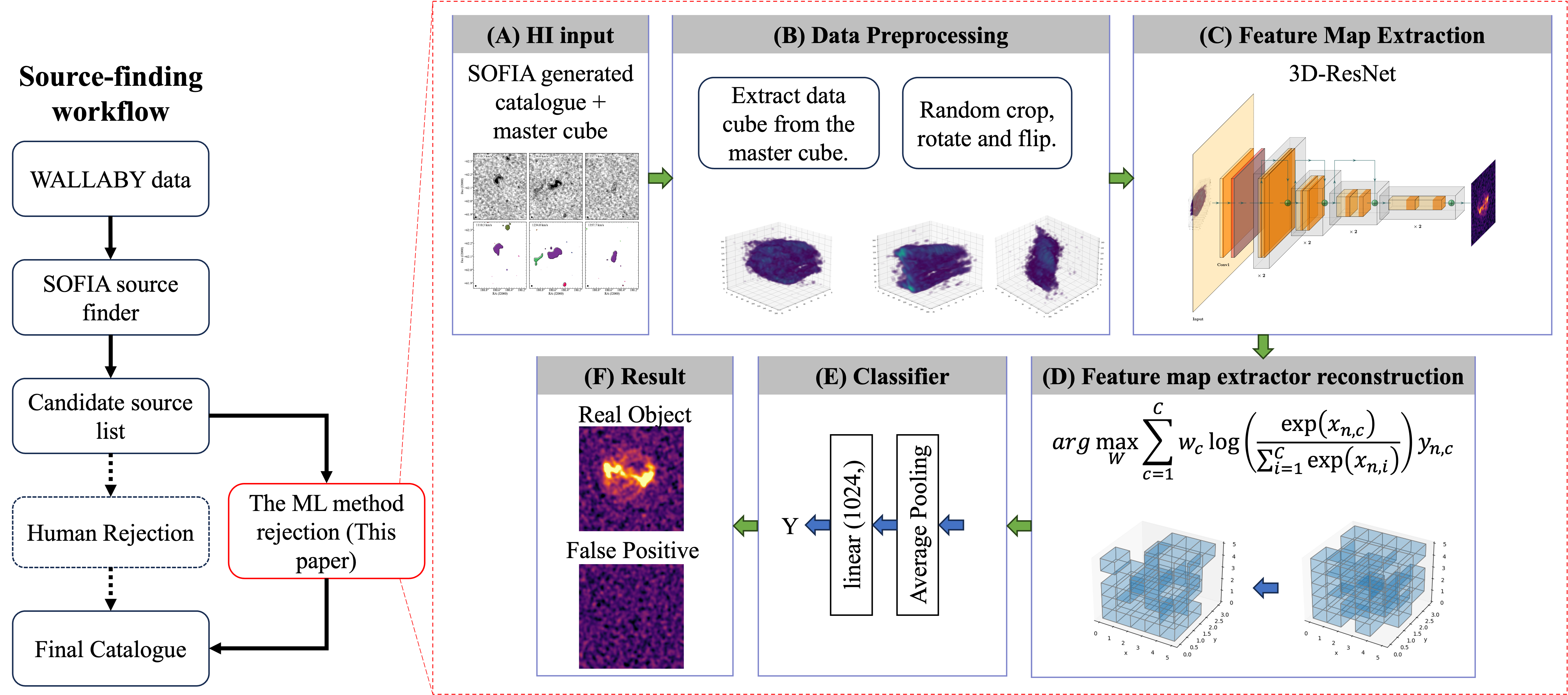}
\caption{Integration of Machine Learning into SoFiA Workflow. On the left, the diagram depicts the comprehensive workflow of SoFiA, within which the right segment illustrates our integrated machine learning approach. The right-hand section details the machine learning pipeline, starting from the HI Input derived from SoFiA's process, proceeding through Data Preprocessing, detailing the feature map extraction strategy, outlining the Optimization Objective, showcasing the Classifier stage, and culminating in the Output Results. This visualization demonstrates how our machine learning methodology fits into and enhances the existing SoFiA workflow.}
\label{fig:overview}
\end{figure*}

{In radio astronomy, convolutional neural networks (CNNs) have been used to classify galaxies based on optical and infrared imaging} \citep{aniyan17,wu19,gupta2023radiogalaxynet,cornu2024yolo}. However, it is important to note that  these applications in astronomy have predominantly focused on 2D image processing of radio continuum observations. In \HI\ surveys, the datasets are almost always three-dimensional, capturing both two-dimensional spatial and one-dimensional spectral, and a new approach is necessary. The third dimension in radio astronomical images provides critical spectral information, adding a layer of complexity to the analysis \citep{tolley2022lightweight}. Recently, \citet{barkai2022} found that SoFiA (in combination with the random forest algorithm) outperforms their V-Net network (plus random forest) \HI\ source finder. 
Recent attempts to apply machine learning to the entire source-finding process in radio astronomy, such as using 3D U-Net for detection by \citet{haakansson2023utilization}, have encountered challenges at the bright end of the flux range. This issue primarily arises from a scarcity of galaxies within that specific region of the parameter space in the training dataset.
However, results from the SKA Data Challenge 2 indicate that integrating traditional source finders like SoFiA with machine learning can enhance \HI\ source-finding performance by up to 20 percent \citep{hartley23}, highlighting the potential of combining conventional and machine learning approaches for improved outcomes in \HI\ surveys.

In this paper, we present a companion machine learning-based model that is more effective at differentiating between the false positives and the true \HI\ detections from SoFiA2's output candidate catalogues. We posit that the implementation of our proposed complementary model will improve upon the efficiency of source-finding in large \HI\ surveys, as the number of false positives will be reduced significantly. 
{Our proposed model employs a three-dimensional (3D) CNN to fully leverage the original 3D data, significantly enhancing the detection and characterization of astronomical sources by exploiting the correlation of true \HI\ emission in the spectral dimension.}

The outline of this paper is as follows. Section~2 provides an overview of our method and workflow. In Section~3, we test our proposed workflow by training and testing on a simulated dataset. This ensures that we are able to quantify the efficacy of our method prior to applying our workflow to ASKAP WALLABY observations. We describe the application of our method to ASKAP data cubes in Section~4.   We then discuss the limitations and implications of our work in Section 5.  Section 6 presents our conclusions and summarises our key results.

\section{Machine learning-based workflow}\label{sec:method}
Machine learning has proven to be exceptionally adept at handling image processing tasks, with architectures like graph neural networks \citep{wang2022pruning}, Residual neural network (ResNet, \cite{he2016deep}) and Transformer models \citep{chen2022auto} showcasing remarkable success in complex visual recognition challenges. The versatility of machine learning extends beyond image analysis, with widespread applications across diverse fields such as environmental science \citep{wang2023machine} and medical diagnostics \citep{chen2019med3d}, showcasing its versatility and effectiveness in interpreting complex datasets.

In this section, we outline our methodology, encompassing data preprocessing, neural network architecture, and training techniques. We detail how we prepare and optimize our dataset, describe our model's structure and layers, and discuss our training strategy, focusing on loss functions, optimization, and overfitting prevention.

\subsection{Pipeline Overview}

Our approach represents a crucial step in the search for \HI\ sources. While utilizing SoFiA, a highly modular and automated tool, proves effective in filtering out the majority of noise, the output from SoFiA still necessitates scrutiny by astronomers to discern genuine sources from processing artifacts or other forms of noise systematics. This is where our machine learning method comes into play, serving to recognize and categorize outputs from SoFiA as either true \HI\ sources, or not.

As illustrated in Fig~\ref{fig:overview}, our machine learning model is designed to supplement and potentially replace the manual inspection phase in the SoFiA workflow, particularly during the initial source list evaluation. By automating this aspect of the process, our method not only streamlines the workflow but also significantly reduces the potential for human error and bias (which are often difficult to quantify). 

\subsection{Pre-processing}

{In this section, we describe the preparation of the data used to train and test our model.
}

\subsubsection{DBSCAN clustering}\label{sec:dbscan}
In our data cleaning process, we addressed the challenge of closely spaced two-dimensional coordinate points in astronomical data, which often represent the same celestial object. Utilizing the Density-Based Spatial Clustering of Applications with Noise (DBSCAN) algorithm \citep{ester1996density}, we identified and excluded sources within a 30 arcsecond radius from each other (consistent with the synthesised beam of the ASKAP WALLABY image cubes). This crucial step of removing redundant data is important for our subsequent analysis.

\begin{table*}[htbp]
\begin{tabular}{llllllll}
\hline
Layer & Function          & Kernel Size & Input size  & Output size & Stride  & Activation & No. of params \\ \hline
1     & Conv3d\_0         & 7, 7, 7     & 1,40,40,70  & 64,40,20,35 & 1, 2, 2 & ReLu       & 21.95 k       \\
2     & maxpool           & -           & 64,40,20,35 & 64,20,10,18 & -       & -          & -             \\
3     & Conv3d\_1.0.1     & 3, 3, 3     & 64,20,10,18 & 64,20,10,18 & 1, 1, 1 & ReLu       & 110.59 k      \\
4     & Conv3d\_1.0.2     & 3, 3, 3     & 64,20,10,18 & 64,20,10,18 & 1, 1, 1 & ReLu       & 110.59 k      \\
5     & Conv3d\_1.1.1     & 3, 3, 3     & 64,20,10,18 & 64,20,10,18 & 1, 1, 1 & ReLu       & 110.59 k      \\
6     & Conv3d\_1.1.2     & 3, 3, 3     & 64,20,10,18 & 64,20,10,18 & 1, 1, 1 & ReLu       & 110.59 k      \\
7     & Conv3d\_2.0.1     & 3, 3, 3     & 64,20,10,18 & 128,10,5,9  & 2, 2, 2 & ReLu       & 221.18 k      \\
8     & Conv3d\_2.0.2     & 3, 3, 3     & 128,10,5,19 & 128,10,5,9  & 1, 1, 1 & ReLu       & 442.37 k      \\
9     & Conv3d\_2.1.1     & 3, 3, 3     & 128,10,5,9  & 128,10,5,9  & 1, 1, 1 & ReLu       & 442.37 k      \\
10    & Conv3d\_2.1.2     & 3, 3, 3     & 128,10,5,9  & 128,10,5,9  & 1, 1, 1 & ReLu       & 442.37 k      \\
11    & Conv3d\_3.0.1     & 3, 3, 3     & 128,10,5,9  & 256,5,3,5   & 2, 2, 2 & ReLu       & 884.74 k      \\
12    & Conv3d\_3.0.2     & 3, 3, 3     & 256,5,3,5   & 256,5,3,5   & 1, 1, 1 & ReLu       & 1.77 M        \\
13    & Conv3d\_3.1.1     & 3, 3, 3     & 256,5,3,5   & 256,5,3,5   & 1, 1, 1 & ReLu       & 1.77 M        \\
14    & Conv3d\_3.1.2     & 3, 3, 3     & 256,5,3,5   & 256,5,3,5   & 1, 1, 1 & ReLu       & 1.77 M        \\
15    & Conv3d\_4.0.1     & 3, 3, 3     & 256,5,3,5   & 512,3,2,3   & 2, 2, 2 & ReLu       & 3.54 M        \\
16    & Conv3d\_4.0.2     & 3, 3, 3     & 512,3,2,3   & 512,3,2,3   & 1, 1, 1 & ReLu       & 7.08 M        \\
17    & Conv3d\_4.0.1     & 3, 3, 3     & 512,3,2,3   & 512,3,2,3   & 1, 1, 1 & ReLu       & 7.08 M        \\
18    & Conv3d\_4.0.1     & 3, 3, 3     & 512,3,2,3   & 512,3,2,3   & 1, 1, 1 & ReLu       & 7.08 M        \\
19    & AdaptiveAvePool3d & -           & 512,3,2,3   & 512,1,1,1   & -       & -          & -             \\
20    & Linear            & -           & 1,512       & 1,2         & -       & -          & 1.03 k        \\ \hline
\end{tabular}
\caption{The network architecture of the 3D ResNet model used in this work. Each convolutional layer is followed by batch normalization and ReLU. Downsampling
is performed by conv3\_1, conv4\_1, conv5\_1 with a stride of 2.}
\label{tb:net_arch}
\end{table*}

\subsubsection{Data augmentation}
To increase the sample size and diversity, we perform standard data augmentation processes such as random cropping, rotation, flipping and resizing. Data augmentation is important for developing robust models that can recognise sources which are not necessarily centred or symmetric within the input training data. {In our study, we employed various data augmentation techniques to enhance the robustness of our model. These techniques included rotation, scaling, flipping, and noise addition, which are well-documented in the literature as effective methods for improving model generalization \citep{shorten2019survey}.}

\subsubsection{Normalization}

Similarly, we implemented common techniques to scale and transform our dataset into a format more suitable for neural network processing. This process is critical to avoid potential biases or misinterpretations caused by the varying scales of raw data values. We employed min-max normalization, which rescales the data into a fixed range of 0 to 1. This approach ensures that each feature contributes proportionately to the final analysis, preventing any single feature from dominating due to scale differences. By normalizing the data, we enhance the efficiency and stability of the neural network's learning process, as normalized data typically result in faster convergence during training \citep{santurkar2018does}. Additionally, this step reduces the complexity of the model's underlying structure, making it less susceptible to overfitting and improving its generalization capabilities on new, unseen data. This normalization approach aligns with standard practices in computer science \citep{singh2020investigating,patro2015normalization} and is pivotal in ensuring that our neural network operates on a consistent and standardized dataset, thereby enhancing the robustness and reliability of our findings.

\begin{figure}
\centering
\includegraphics[width=0.75\linewidth]{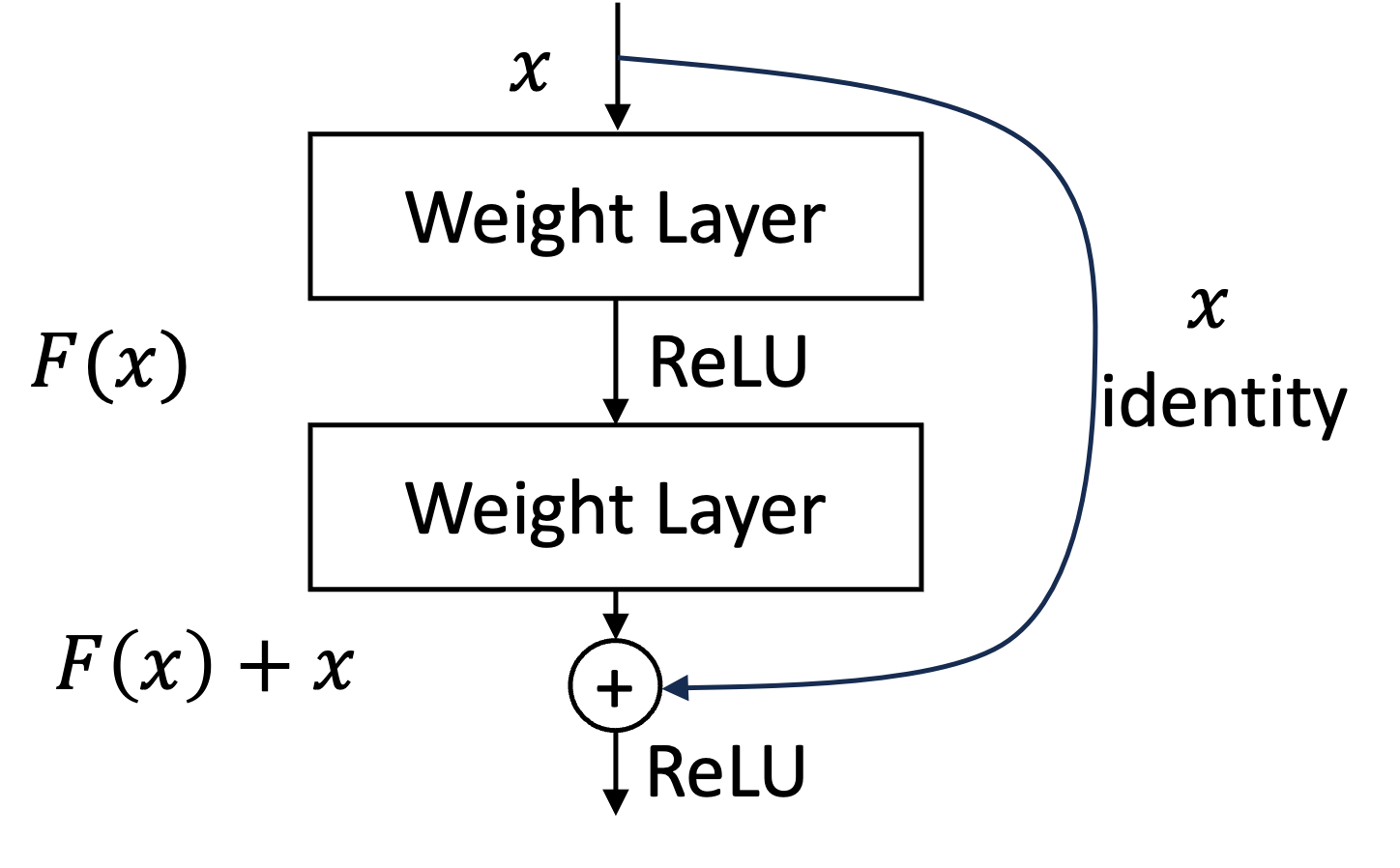}
\caption{Residual block. Shortcut connections bypass a signal from the top of the block to the tail. Signals are summed at the tail.}
\label{fig:resnetblock}
\end{figure}

\subsection{3D convolutional network}
{A Convolutional Neural Network (CNN) \citep{o2015introduction} is a type of deep learning model particularly effective for processing grid-like data, such as images. It consists of layers of neurons that use convolutional kernels to detect features within the data. During training, the network adjusts its kernel parameters to minimize the error in predicting the correct labels for the data.}
Our architecture utilizes ResNets \citep{he2016deep} as its foundation. ResNets incorporate shortcut connections, allowing a signal to skip directly from one layer to another. These connections enable gradients to flow more effectively from later layers back to earlier ones, simplifying the training of particularly deep networks. Figure \ref{fig:resnetblock} illustrates the residual block, a fundamental component of ResNets. In this block, signals are directly channeled from the beginning to the end. ResNets are composed of numerous such residual blocks.

Table \ref{tb:net_arch} describes the architecture of the 3D ResNet network\citep{fan2017lung}. The primary distinction between our version and the original ResNets lies in the dimensions of the convolutional kernels and pooling operations. Our 3D ResNets utilize 3D convolution and 3D pooling. The 3D convolutional layer can be described as the output value of the layer with input size $(N, C_{in}, D, H, W)$.  The output \\ $(N, C_{out}, D_{out}, H_{out}, W_{out})$ is described as:
\begin{equation}
    \textrm{out}(N_i,{C_{out}}_j)=\textrm{bias}(C_{out_j}) + \sum^{C_{in}-1}_{k=0}\textrm{weight}(C_{out_j},k)*\textrm{input}(N_i,k),
\end{equation}
where $*$ is the valid 3D cross-correlation operator, {and $N$ is a batch size, $C$ stands for the number of channels, $D,H,W$ is the depth, height, and width of input planes, respectively. }
The convolutional kernels measure [3, 3, 3], and the temporal stride for the 'Conv3d\_0' layer is set at 1. The network processes input cubes with dimensions of [1, 70, 40, 40]. The sizes of these input clips are determined by the median value from the cube size statistics output by SoFiA. Down-sampling is executed by layers 'Conv3d\_2.0.1', 'Conv3d\_3.0.1', \\and 'Conv3d\_4.0.1', each using a stride of 2. When the number of feature maps escalates, we implement identity shortcuts combined with zero-padding to prevent an increase in parameter count.

The output of each block serves as the input for the subsequent block. This stacking mechanism is crucial, as it augments the number of non-linear activations. Each convolutional layer comes equipped with its own Rectified Linear Unit (ReLU) \footnote{The ReLU activation function is widely used in computer vision and deep learning for more effective training.} \citep{agarap2018deep} which integrates non-linearities into the system.   
{These non-linear activations enable the network to model complex patterns and relationships within the data, thereby enhancing its ability to extract distinctive features. In the context of neural networks, a neuron refers to a computational unit that receives input, processes it through a non-linear activation function, and passes the result to the next layer \citep{lecun2015deep}. }
It's worth highlighting that the receptive field size of an individual neuron doesn't restrict our proposed method from identifying sources that are more expansive. This is attributed to the fact that {a feature map is an aggregation of several neurons}, which, when combined, have the capacity to detect considerably larger entities.

\begin{figure*}[ht]
\centering
\includegraphics[width=0.7\linewidth]{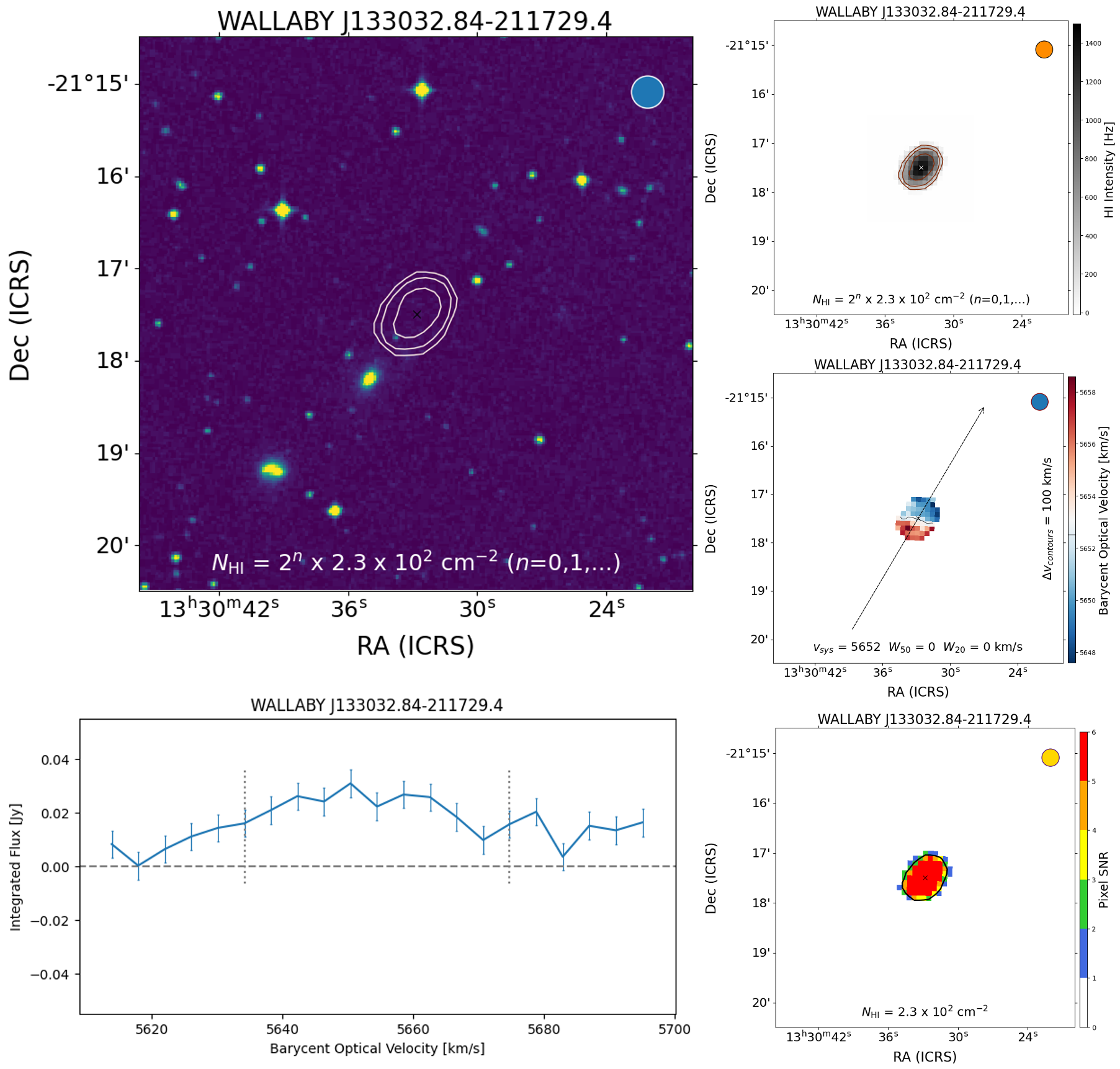}
\caption{Based on the input WALLABY image WALLABY J133032-211729, examples of derived data sets. The panels are HI contours overlaid on optical image (Top left), HI contours overlaid on multiwavelength image (top right), velocity map showing the galaxy rotation (middle right), pixel-by-pixel SNR maps (bottom right), spectra without noise (bottom left).}
\label{fig:WALLABY_image}
\end{figure*}

Figure~\ref{fig:overview}(D) presents the feature maps generated by the concluding convolutional layer, specifically, 'layer18 \\ Conv3d\_4.0.1' as referenced in Table \ref{tb:net_arch}. These features are derived from the input cube labeled "\textcolor{black}{WALLABY J133032-211729}" shown in Figure~\ref{fig:WALLABY_image}.
{Our model outputs a 3D feature map, effectively capturing the spectral features of the data.}
The feature map is updated after comprising 20,000 iterations of forward computation paired with backward propagation. This rigorous process was essential to pinpoint the optimal values for all kernel weights within the model. Detailed insights into the training and optimization phases are described in the next section. A cursory visual evaluation indicates a noticeable similarity between the original input image and each of the feature maps, particularly in terms of the source morphology.  Every individual feature map unveils unique attributes, each a product of a specific kernel set. Each kernel within this set has been trained to discern and align with a distinct pattern from its respective input tensors.  

The efficiency and effectiveness of the training pipeline are largely determined by the training loss, which can be expressed as follows:
\begin{equation}
    \textrm{Loss}=-\sum^C_{i=1}y_i\cdot \textrm{log}(\hat{y}_i),
\end{equation}
where $C$ is the total number of classes, $y$ is the one-hot encoded vector\footnote{a binary vector with all zero values except for a single one at the position corresponding to the class.
}, and $\hat{y}$ is the model's predicted output. The goal during training is to reduce the training error on the training set using various optimization techniques without compromising the model generality on future unseen datasets.

\section{Verifying our workflow}
We first validate our machine learning-based approach using simulated data which ensures that we have an excellent `ground truth' dataset down to low SNR --- not typically available in real datasets. This section outlines the creation of our simulated dataset, the implementation details, and a concise analysis of the outcomes. 

\subsection{Dataset Generation}
To evaluate our model's effectiveness regarding completeness, and reliability, we produced {4,000} simulated galaxies utilizing the GALMOD function in the GIPSY data processing software \citep{allen2011gipsy}. We varied several galaxy parameters randomly within a reasonable range to ensure a diverse array of observational characteristics. These parameters included peak HI column density (10$^{20}-$10$^{21}$ cm$^{-2}$), rotation velocity (ranging 30 to 220 \kmps), scale length on the sky (4.5 to 36 arcseconds), disc inclination (0 to 85 degrees), and position angle (0 to 360 degrees). Consistent with the WALLABY restoring beam, these model galaxies were convolved with a 30~arcsecond Gaussian beam. Each galaxy's flux density was uniformly adjusted to ensure most integrated SNRs would lie in the 0–-10 range.

{
We also generated a corresponding number of negative samples, totaling 4,000. 
These negative cubes were randomly sampled from a master data cube featuring authentic noise from ASKAP WALLABY observations. The master data cube was produced following a procedure similar to that detailed in \citep{westmeier2021SoFiA}. }
We extracted the noise from a 1501 × 1501 spatial pixel and 1501 spectral channel section of a WALLABY pre-pilot data cube from the Eridanus cluster pointing, creating a file around 12.6 GB in size. The simulated cube has pixel sizes of 6 arcseconds (synthesised beam of 30 arcseconds) and a spectral channel width of 18.5~kHz, translating to a velocity resolution of about 4 \kmps at a redshift of 0. To reduce the likelihood of contamination from actual \HI\ emissions, the noise cube was sourced from the 1323--1351~MHz frequency range, where very few HI sources are found.
{These positive and negative samples directly constituted the model's dataset and were not further processed by SoFiA.}

\subsection{Implementation and evaluation}\label{sec:metric}
We implement the method using PyTorch \citep{paszke2019pytorch}. Both training and testing require GPU resources, and we deploy the model to run on NVIDIA RTX A5000 (16GB RAM) GPU.
To train our network, we employ stochastic gradient descent (SGD) combined with momentum \citep{ruder2016overview}, and set the initial learning rate to 0.001. The training speed is about 0.03 s per iteration on A5000. Thus, a pipeline instructed to execute 20,000 iterations requires 600~s of training time on provisioned GPU resources. 
For testing, it takes the learned model 45--220 milliseconds per subject to generate detected radio sources and probabilities. As is typical, this time cost is highly dependent on I/O performance. 

{Here, we divided the simulated dataset into training, validation, and test subsets in a 0.8, 0.15, and 0.05 ratio.} This split was strategically chosen to ensure ample data for comprehensive model training while maintaining separate, untouched datasets for validation and unbiased testing. Allocating 80\% to training provides the model with extensive learning opportunities. The 15\% validation set enables effective tuning and overfitting prevention, and the 5\% test set ensures the model's performance is evaluated on completely new data, reflecting its real-world applicability and accuracy. This approach ensures a balanced and rigorous assessment of the model's capabilities.

To evaluate the proposed method against the testing set, we use the evaluation metrics of accuracy and the F1 score. {Accuracy represents the fraction of classifications that are correct, as shown in Eq. \eqref{eq:acc}.} Precision and completeness are defined in Eq. \eqref{eq:precision} as follows:
\red{\begin{equation}\label{eq:acc}
\text{Accuracy} = \frac{\text{TP} + \text{TN}}{\text{TP} + \text{TN} +\text{FP} + \text{FN}}
\end{equation}
\begin{equation}\label{eq:precision}
            \text{Precision}=\frac{\text{TP}}{\text{TP+FP}} \; \text{and} \; \text{Completeness} = \frac{\text{TP}}{\text{TP+FN}},
\end{equation}}
where true positive (TP) is the number of items correctly identified as true, false positives (FP) is the number of items incorrectly identified as true, and false negatives (FN) is the number of items incorrectly identified as false.
Then, the F1 score is defined as:
\red{\begin{equation}\label{eq:f1}
    \text{F1}=2\times \frac{\text{2TP}}{\text{2TP} + \text{FP} + \text{FN}}
    \end{equation}}

\subsection{Results}
The results from the mock galaxy dataset, obtained using three different random seed runs, demonstrate the robustness and reliability of our model under varying conditions. As shown in Table~\ref{tab:num_results}, we find high test accuracy, averaging $\sim 96\%$ across all three runs, indicating a strong performance in source identification. Furthermore, the low variance associated with this accuracy, despite the different random seeds, underscores the model's stability and predictability. The validation F1 Score of $\sim 96\%$ further confirms the model's balanced precision and completeness, a critical aspect in astronomical data analysis where imbalanced classes may influence the accuracy scores.  In summary, our experiment with the simulated dataset verifies our model's ability to deliver consistent and reliable outcomes.

\begin{table}[ht]
\centering
\caption{Performance Metrics of Our Method on Mock Galaxy Dataset}\label{tab:num_results}
\begin{tabular}{|l|l|l|}
\hline
\textbf{Metric}         & \textbf{Mean (\%)}    & \textbf{Variance} \\ \hline
Best Test Accuracy      & 95.82           & 0.2291        \\ \hline
Best Train Accuracy     & 98.67           & 0.1846        \\ \hline
Test F1 Score     & 95.71           & 0.2248        \\ \hline
\end{tabular}
\end{table}

\begin{figure}
    \centering
    \includegraphics[width=0.75\linewidth]{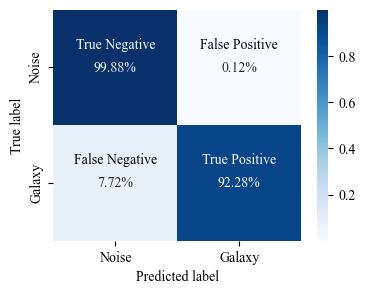}
    \caption{The confusion matrix illustrates the model's performance in classifying data as either 'Galaxy' or 'Noise.'}
    \label{fig:cmatrixsim}
\end{figure}

The confusion matrix (Figure~\ref{fig:cmatrixsim}) generated from the model's predictions demonstrates excellent performance in differentiating between simulated galaxies from noise. We find a high True Positive Rate (TPR) of 92.28\% and a high True Negative Rate (TNR) of 99.88\%, suggesting that our model is reasonably complete and reliable. On the other hand, the False Positive Rate (FPR) and the False Negative Rate (FNR) are at 0.12\% and 7.72\%, respectively. The very low FPR is reassuring but the FNR suggests that a small percentage of true \HI\ detections have been missed by the model, lowering our method's accuracy.

\begin{figure}
    \centering
    \includegraphics[width=0.95\linewidth]{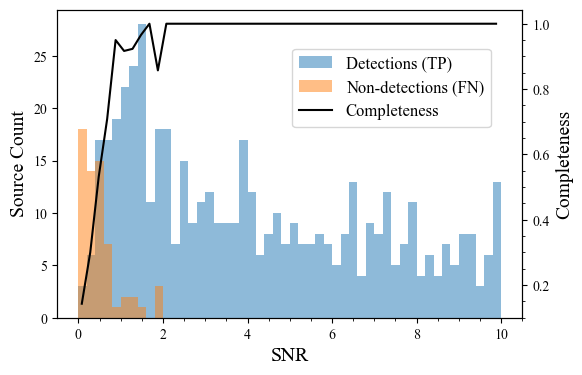}
    \caption{Histogram of detected (blue) and undetected (orange) mock galaxies and the completeness (black) as a function of SNR, demonstrating that the model is able to achieve 100 percent completeness at SNR$\gtrsim$2}
    \label{fig:SNR}
\end{figure}

In order to further analyse the completeness of the source finding run as a function of SNR, we establish a way of characterizing the integrated SNR of a source in the same fashion as \citet{westmeier2021SoFiA}. {Specifically, the SNR is calculated by taking the ratio of the peak signal intensity of the source to the standard deviation of the noise in the data.} From the injection of simulated \HI\ sources into real ASKAP WALLABY noise cubes, we are able to compare the relationship between completeness and SNR. 
While astronomers often find visual identification of \HI\ sources at low SNR (SNR$\approx$3--5) to be challenging (and often rely on additional multiwavelength information to make more accurate judgements), we find the model generated by our machine learning workflow to be remarkably accurate, even down to low SNRs of $\sim2-3$ (Figure~\ref{fig:SNR}).  Therefore we have demonstrated here that our method is able to generate a quantifiably reliable catalogue of true sources from the extensive SoFiA candidate catalogues.

\section{Application to WALLABY DR2 pilot data}
In this section we assess our model's performance on real \HI\ sources from the ASKAP WALLABY DR2 pilot observations (Murugeshan et al in prep.) observations.
This direct comparison ensures consistency in performance evaluation, as we analyze a dataset of 11,121 candidate sources. 
The section delves into the { dataset specifics}, evaluates the model quantitatively, and includes a visual review of the findings. Through this approach, we verify our model's capability to effectively interpret and work with real observations, and if our model's robustness on real observations is comparable to that of the simulated datasets.

\subsection{{Dataset preparation for ASKAP observations}} \label{sec:dirty}

In order to ensure no potential galaxies are overlooked, the SoFiA algorithm is tuned to operate with a high degree of sensitivity. {We refer to the outputs from SoFiA, which include a large number of false positive objects, as candidate sources.} However, this increased sensitivity results in generating a significant number of false positives.  Taking the NGC~5044 pointings (from DR2) as an example, SoFiA identifies 11,121 candidate sources. {Despite efforts to manually adjust parameters and other methods to eliminate false positives, these adjustments are relatively biased towards retaining all possible objects to minimize the risk of missing actual objects. Yet, upon further analysis by astronomers, only 1,326 (11.92\%) of these sources have been confirmed as actual galaxies.}

{We begin with the candidate catalogue from SoFiA, which contains 11,121 sources. To clean the dataset, we use the Density-Based Spatial Clustering of Applications with Noise (DBSCAN) algorithm as described in Section \ref{sec:dbscan}. Please note that DBSCAN is used solely for data cleaning and is not a part of our model.}

\begin{figure}
    \centering
    \includegraphics[width=0.95\linewidth]{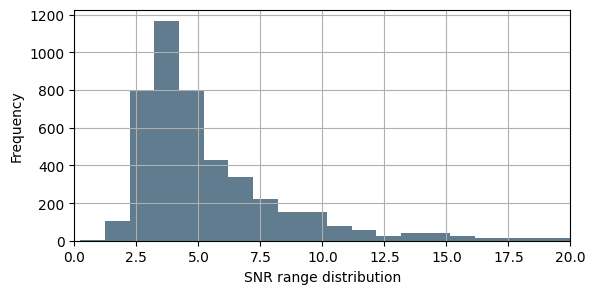}
    \caption{The distribution of the SNR in the data set that consists of 5,889 potential subjects selected from DR2. }
    \label{fig:datasetSNR}
\end{figure}

Upon applying the cleaning criteria, we obtain a dataset $\mathcal{D}$ that has 5889 HI sources. We show the SNR distribution of these 5889 \HI\ sources in Figure~\ref{fig:datasetSNR}. 
We find a unimodal distribution centered at SNR$\approx$4 and an asymmetric tail that is extended towards higher SNR, indicating that brighter sources are rarer. The majority of sources have SNRs between 2.5 and 7.5, with fewer having SNRs greater than 10.  
The input layer dimension in our model was configured to be [40, 40, 70], a decision informed by comprehensive data analysis. 
{The input layer size is determined based on the 95th percentile of the spatial data distribution and 90th percentile of the spectral data distribution, meaning that 95\% or 90\% of the data is smaller than this size. In practice, due to preprocessing, cubes that are larger or smaller than this size are resized using interpolation to fit this size.} 
We random split this dataset into 3 subsets {(equal portions of positive and negative)} for the training set ($80\%$), the validation set ($15\%$) and the test set ($5\%$). {The model was not exposed to the test set before testing, adhering to standard machine learning practices.}

\subsection{Results of our model on WALLABY DR2 pilot data}

\begin{figure}
    \centering
    \includegraphics[width=0.8\linewidth]{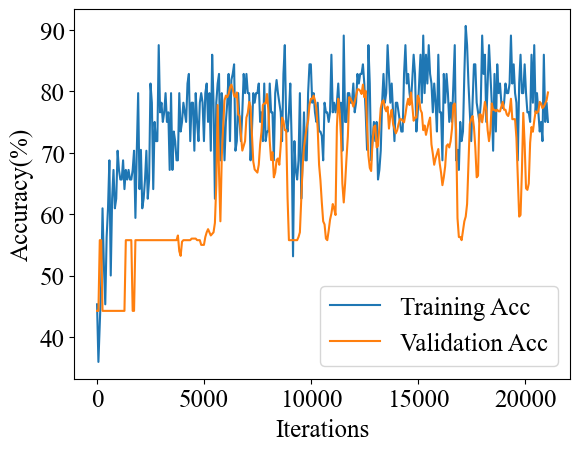}
    \caption{Learning curves monitor the change of training (blue curve) and validation (orange curve) accuracies (Y-axis) as the training progresses by number of iterations (X-axis) .}
    \label{fig:loss}
\end{figure}

We employed both Adam and SGD optimizers to perform gradient descent on the neural network, aiming to improve the training accuracy on the training dataset while preserving the model's generalization capability on unseen datasets. To illustrate the variation in loss during the training process, we plotted the training curves in Fig. \ref{fig:loss}, where the Y-axis represents model accuracy training set and the X-axis indicates training iterations. As the training progresses, the training accuracy increases progressively, rising from 0.45 to 0.80. The accuracy exhibits a rapid increase during the first 7,000 iterations, followed by a more gradual improvement. After approximately 20,000 iterations, the upward trend in both curves begins to plateau, suggesting that the model has attained its accuracy limit given the existing network configuration and dataset. To prevent overfitting, we employed the early stopping technique \citep{smale2007learning} to halt the training process.

To determine the optimal ResNet architecture for our specific purpose, we test and compare the performance of three ResNet architectures: ResNet18, ResNet34 and ResNet50.  The different version numbers describe the number of convolutional layers in each of them.   Table \ref{tab:ASKAPresult} compares our performance across all three ResNet versions. In general, without sufficient data, the model may struggle to learn more complex patterns, even as its capacity increases with more layers.

These results highlight ResNet18's efficacy, particularly noteworthy given its computational efficiency relative to more complex models.
These quantitative outcomes indicate that the computationally lighter ResNet18 model is not only capable of providing high accuracy in distinguishing true galaxies from artefacts in SoFiA data but does so with a consistency that rivals or exceeds that of its more complex counterparts. This suggests that for tasks requiring the identification of \HI\ sources where computational resources may be a constraint, ResNet18 offers a balanced solution between performance and resource utilization.

\begin{table}[htbp]
\begin{tabular}{llll}
\hline
Method    & Resnet18 & Resnet34 & Resnet50 \\ \hline
Train Acc & 	79.77$\pm$0.15  & 77.50 $\pm$1.58         &   80.80$\pm$1.72       \\ \hline
Val Acc   &  76.73 $\pm$ 2.91         &   75.78 $\pm$4.20       &  75.62 $\pm$ 1.91       \\ \hline
Test Acc   &  76.92 $\pm$ 0.77         &   76.39 $\pm$3.20       &  76.41 $\pm$ 2.22       \\ \hline
Test F1   &  78.92 $\pm$ 1.46        &   77.56 $\pm$4.70       &  78.35 $\pm$ 3.56       \\ \hline
\end{tabular}
\caption{Comparative Performance Metrics of ResNet Architectures on SoFiA Output Data.}
\label{tab:ASKAPresult}
\end{table}

\begin{figure}
    \centering
    \includegraphics[width=0.75\linewidth]{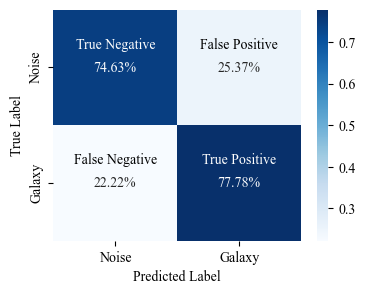}
    \caption{Confusion Matrix showcasing the performance of our model on real astronomical data. The matrix quantifies the model's ability to distinguish between actual galaxies and noise/artifacts, reflecting the real-world complexities such as lower SNR and the presence of artifacts.}
    \label{fig:cmatrixreal}
\end{figure}

Figure~\ref{fig:cmatrixreal} presents the confusion matrix from the application of our model to ASKAP data and presents insightful outcomes. 
This confusion matrix demonstrates that our model is able to correctly identify true positives (actual galaxies) and true negatives (actual noise or artifacts). The high TPR of 77.78\% suggests that the model is effectively identifying a large portion of genuine galaxies in the data. Similarly, the TNR of 74.63\% indicates that the model is proficient at recognizing noise or artifacts, which is crucial in a real-world astronomical setting where noise levels are higher and SNR (Signal-to-Noise Ratio) is lower compared to simulated data. 

The FPR of 25.37\%, though higher than ideal, is a reflection of the challenging nature of real data, which often includes more complex noise patterns and processing (or imaging) artifacts. This may lead to a higher rate of false positives, where noise or artifacts are incorrectly classified as galaxies. 

Similarly, the FNR of 22.22\% indicates that a portion of actual galaxies is being missed. Twelve DR2 sources with SNR$>5$ were missed by our model despite the high SNR nature of these sources.  This suggests that while our model is able to characterise the general properties of true \HI\ detections, there appears to be a greater range of properties possessed by true \HI\ sources (than generalised by our model).  A greater number of sources may be required in the training sample to improve upon the understanding of \HI\ detections with more extreme properties. {We acknowledge that the small test set size may introduce variability due to small number statistics. However, we have tested different training-test splits with various random seeds, and the TPR and TNR remained consistent within a reasonable range. }

\begin{figure}
    \centering
    \includegraphics[width=0.95\linewidth]{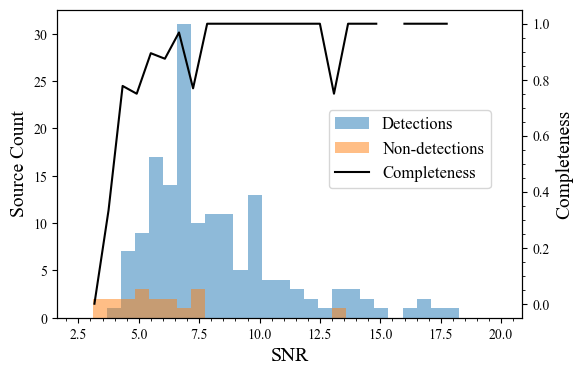}
    \caption{Histogram of detected (blue), undetected (orange) real galaxies, and the completeness (black) as a function of SNR.}
    \label{fig:SNRreal}
\end{figure}

\begin{figure}[ht]
 \centering
 \begin{subfigure}[b]{0.85\textwidth}
    \includegraphics[width=1\linewidth]{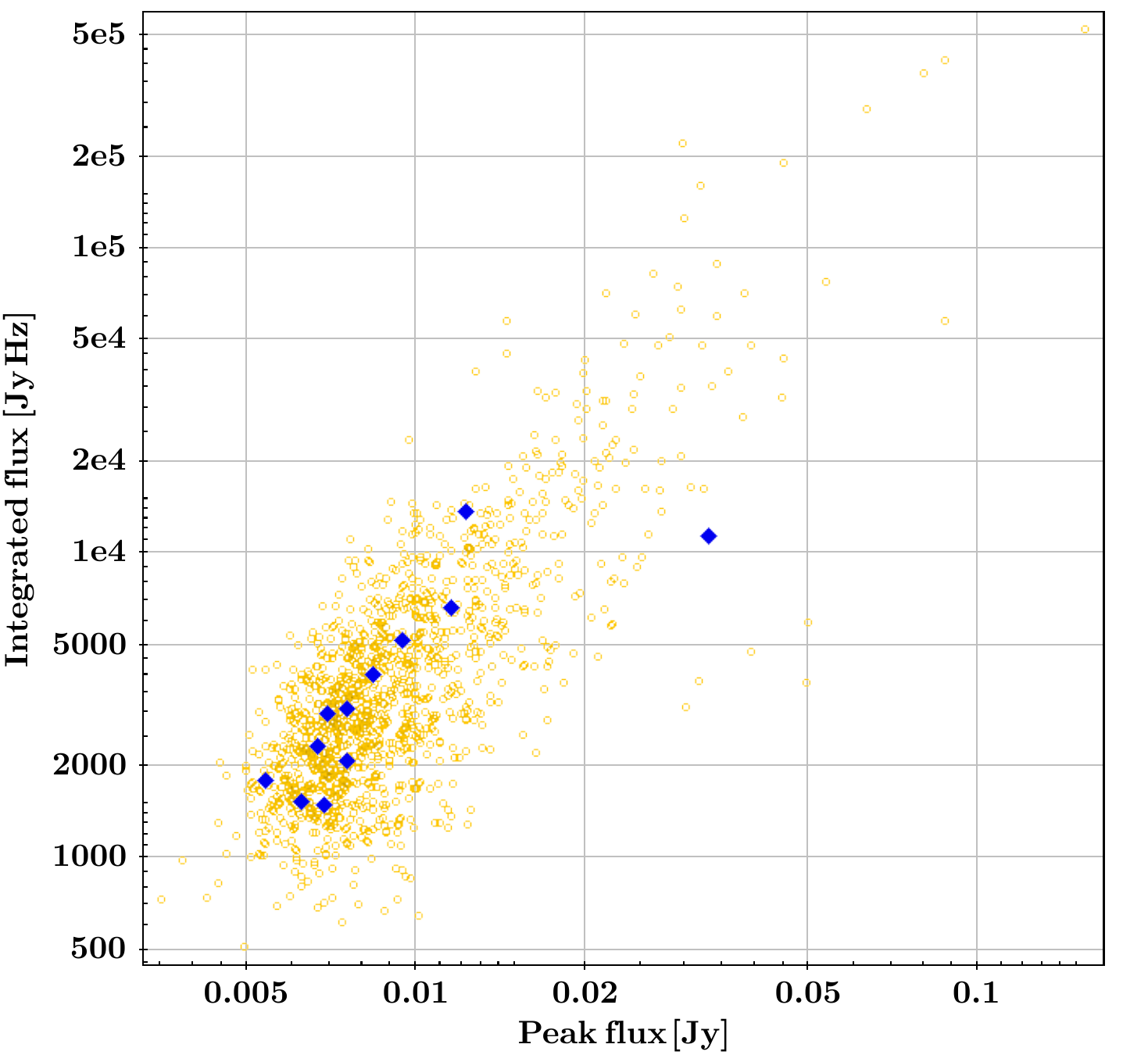}
    \caption{}
    \label{fig:FN-fmaxfsum} 
 \end{subfigure}

 \begin{subfigure}[b]{0.85\textwidth}
    \includegraphics[width=1\linewidth]{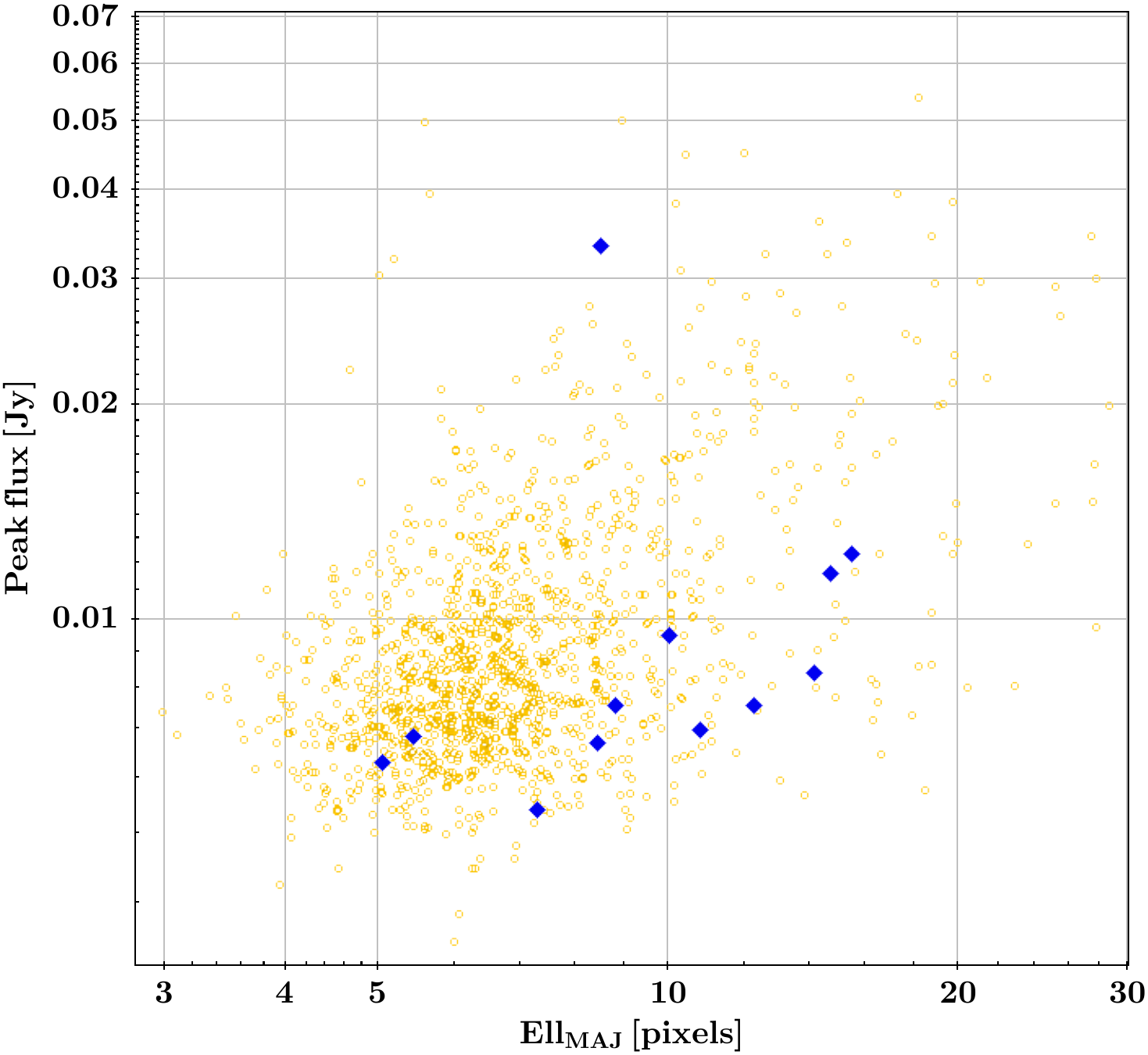}
    \caption{}
    \label{fig:FN-ellmajfmax}
 \end{subfigure}

 \begin{subfigure}[b]{0.85\textwidth}
    \includegraphics[width=1\linewidth]{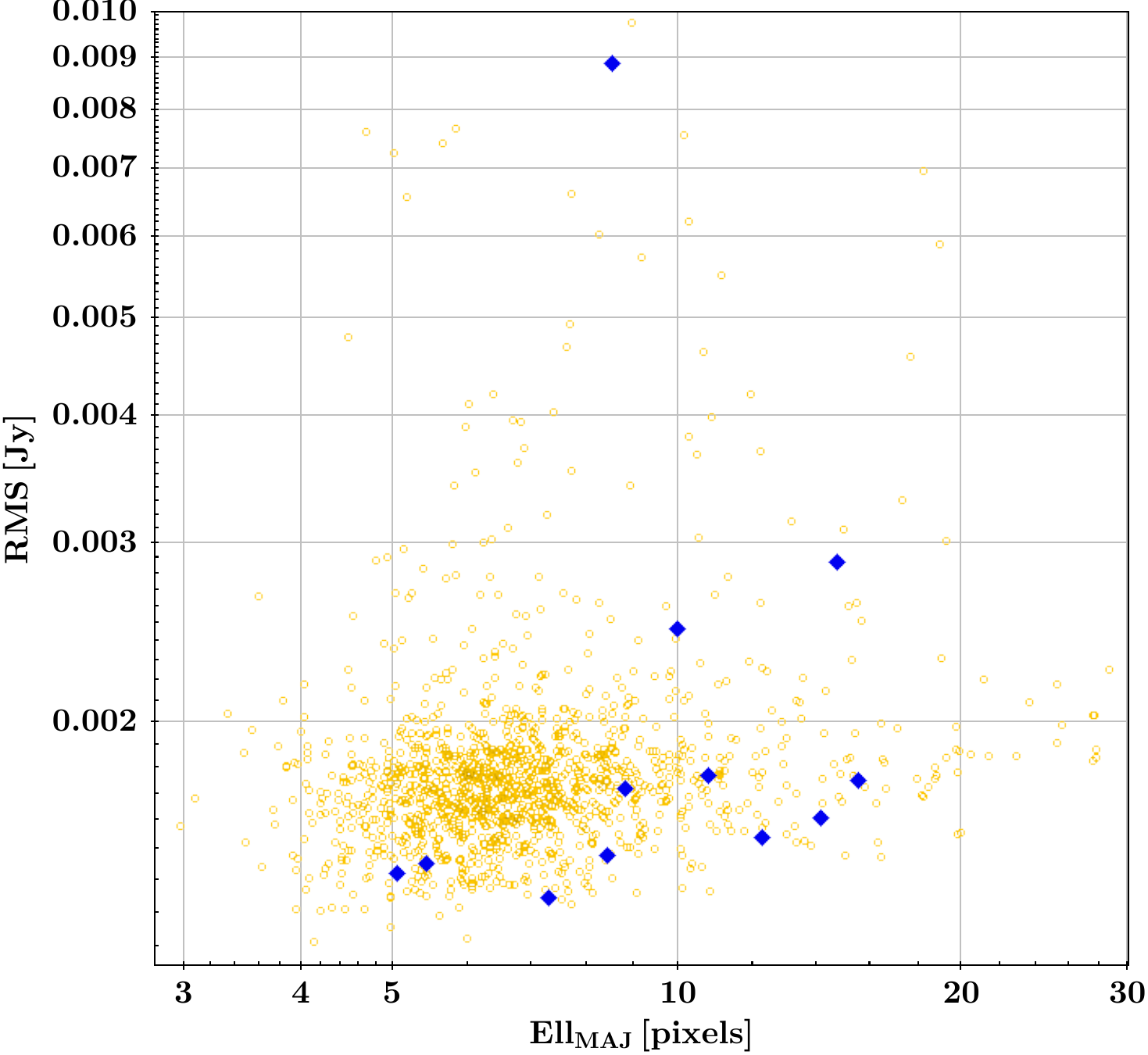}
     \caption{}
    \label{fig:FN-ellmajrms}
 \end{subfigure}

\caption{Panels (a), (b) and (c) show the distribution of integrate flux, peak flux and RMS flux for the DR2 SoFiA candidate list (small yellow open circles) and the twelve false negative sources missed by our model (large blue filled diamonds).}\label{fig:modelnondetect}
 \end{figure}

Visualising the model's accuracy as a function of SNR, we find that our model is able to achieve reliable accuracy at SNR$>$7.5 (Figure \ref{fig:SNRreal}). However, we missed a source with SNR = 11, as shown in Figure \ref{fig:WALLABY_image}. Why is our model not recovering all the 1,326 sources catalogued in the NGC~5044 pointings of DR2? The likely reasons are:

\begin{enumerate}
    \item Inherent bias in the range of SNR in our sample.  Our WALLABY DR2 dataset used for training, validation and testing contains sources which typically have SNR$>$5, limiting our model's ability to perform at lower SNR.  Therefore, we do not expect our model to surpass the performance set by the data on which it was trained.
    \item More extreme or complex properties associated with high SNR sources. As can be seen in Figure~\ref{fig:SNRreal}, our model is also misclassifying high SNR \HI\ sources.  Related to the narrow range of properties described by our DR2 dataset, we hypothesise that we are missing these high SNR sources due to rarer properties that have not been modelled well by our current model.  While the multidimensional feature maps may be more difficult to interpret, we examine the known properties of the \HI\ sources that have been misclassified as FN to illustrate the outlier nature of these FN sources.  Figure~\ref{fig:modelnondetect} shows that while the peak and integrated \HI\ fluxes of FN sources are consistent with that of the general population, we find that our model's FN sources typically reside in outlying parameter spaces relative to the general population in terms of size (as traced by the ellipse major axis, $ell_{\rm{maj}}$); \HI\ line width, $W20$; and noise, $rms$.  A reason for the rarity of some of these sources comes from the observational constraints and limitations. For example, it is typically quite difficult to detect a bona fide \HI\ source that has both large angular extent and wide \HI\ line width.
\end{enumerate}

While our model results on the WALLABY DR2 data may not match the completeness seen for the simulated datasets, they nonetheless demonstrate the model's practical utility in assisting astronomers with source-finding. By effectively reducing the volume of data through the accurate identification of a majority of true galaxies and noise/artifacts, the model can significantly streamline the data analysis process, allowing astronomers to focus on the most promising data for further investigation. This efficiency is particularly valuable in large-scale surveys, where the sheer volume of data can be overwhelming.

\begin{table*}
\caption{Properties of the additional \HI\ sources. Col (1): Name of the source; Col (2): Optical ID of the associated galaxy;  Col (3): Right Ascension (RA) centre of the \HI\ emission; Col (4): Declination (Dec) centre of the \HI\ emission; Col (5): Central \HI\ velocity (in optical convention); Col (6): Integrated \HI\ flux; Col (7): Width of \HI\ emission line at full-width-half-maximum; Col (8): Comments about the source including the optical identity of the source.}
\begin{tabular}{llcccccl}
\hline
Source  & Optical ID &  RA  & Dec & $v$ & $S_{\textrm{INT}}$  & $W_{\rm{50}}$ & Comments  \\
      & & (J2000)  & (J2000) & km~s${^{-1}}$ & Jy~km~s${^{-1}}$ & km~s${^{-1}}$& \\
(1)  & (2)  & (3) &  (4)  & (5) & (6)  & (7) & (8)  \\
\hline
WALLABY J130204-112248 & NGC~4290 &13:02:04 & -11:22:48 & 1334 &360.5 & 93 &  12-arcsecond map available \citep{murugeshan24} \\
WALLABY J131851-210223 & NGC~5068& 13:18:51 & -21:02:23 & 669 & 4409.3 & 67 & 12-arcsecond map available \citep{murugeshan24} \\
WALLABY J133612-221125 & LEDA 817885 & 13:36:12 & -22:11:25 & 8316 & 9.1 & 76 & new \HI\ detection\\
\hline
\end{tabular}
\label{tab:newsource}
\end{table*}

\subsection{Recovery of additional \HI\ sources from the unverified SoFiA2 candidate catalogue}

As an additional test, we apply our model to {previously mentioned 11,121 data cubes in \ref{sec:dirty} identified as candidate sources by SoFiA} for the NGC~5044 data cubes that were observed by ASKAP in 2022. Can our model identify additional \HI\ sources that have not been catalogued via the default WALLABY source-finding process \citep{westmeier22dr1,murugeshan24}?  \red{Please note, this dataset of 11,121 sources is a combination of the training, validation, and test sets.} 

\begin{figure}[!htbp]
    \centering
    \includegraphics[width=0.8\linewidth]{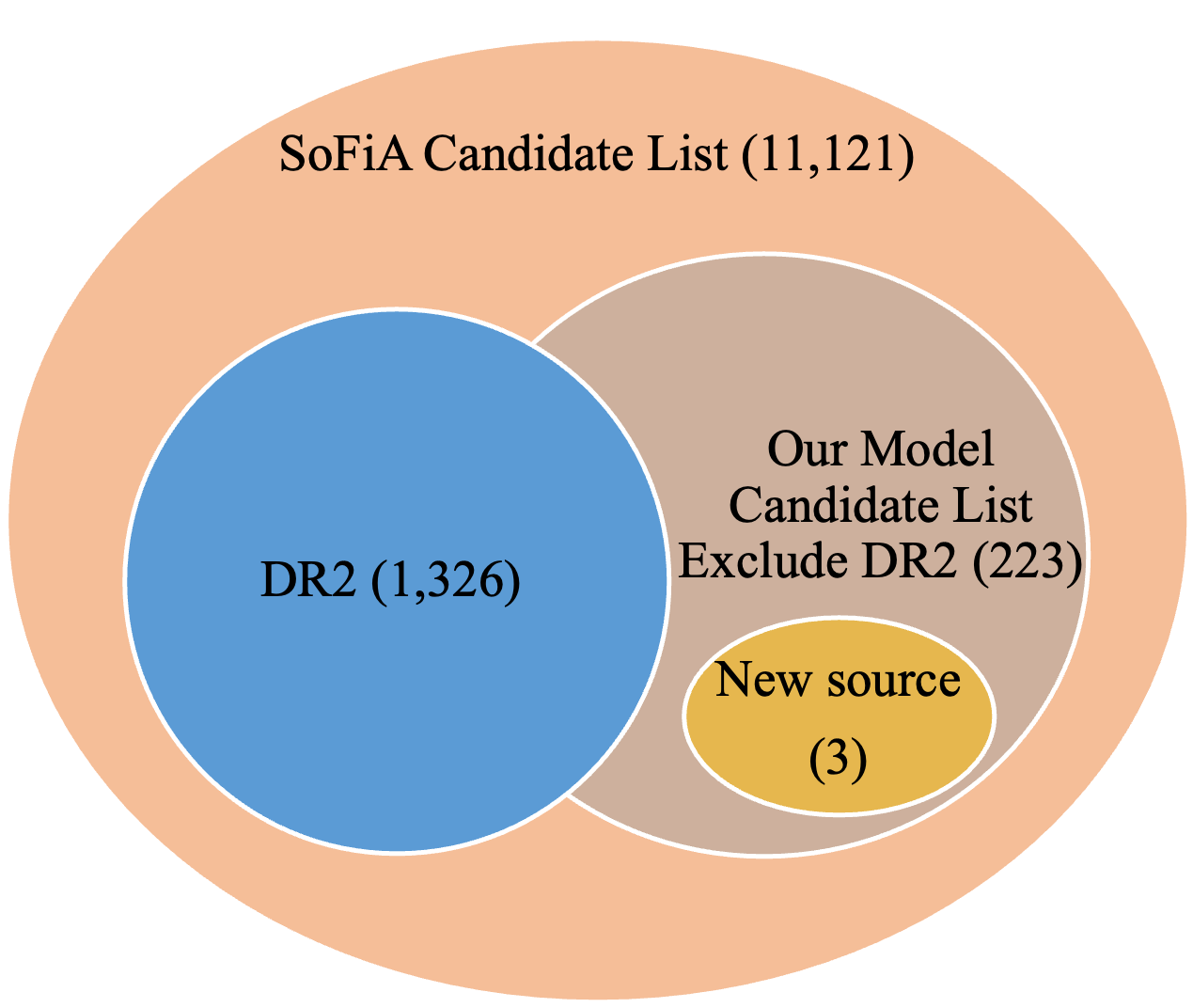}
    \caption{The relationship between the candidate lists and new sources found.}
    \label{fig:venn}
\end{figure}

Similar to SoFiA, our model generates a list of candidate \HI\ sources.  However, this candidate list is a much smaller subset than that of the original SoFiA candidate list. After removing catalogued DR2 sources \citep{murugeshan24}, we were left with a list of {223} candidates. This can be compared with the initial SoFiA candidate list of {11,121} candidates, of which 1,326 \HI\ sources, {as described in Section \ref{sec:dirty}.}
The relationship between the candidate lists are shown in Figure \ref{fig:venn}.  

 In addition, we found three additional \HI\ sources that have not been catalogued in the WALLABY DR2 30-arcsecond catalogue \citep{murugeshan24}.  Two of the three additional sources identified by our model are large extended \HI\ nearby galaxies (NGC~4920 and NGC~5068) that have been presented in \citet{murugeshan24} as part of their high-resolution sample --- a sample of nearby \HI\ sources previously catalogued in the HI Parkes All Sky-Survey \citep{koribalski04,meyer04,wong06}. These two sources were left out of the default 30-arcsecond WALLABY DR2 catalogue due to their position and extent near the edges of their respective source-finding regions.  Hence, these two sources will be recovered in future WALLABY data releases when additional sky regions are observed and these two sources are further away from the edges of their respective fields.

The third additional source is a new \HI\ detection of a more distant galaxy, LEDA~817885.  The \HI\ central velocity is consistent (within uncertainties) with previous spectroscopic measurements of the recessional velocity of LEDA~817885 \citep{jones09}. On the other hand, the \HI\ position centre is slightly offset by approximately 28~arcseconds to the north-east of the galaxy's optical centre.  It appears that the north-eastern region is more \HI-rich than the south-western region of LEDA~817885.  The recovery of this source alone argues for the benefits of source-finding using multiple approaches.

\begin{figure*}[ht]
\centering
\begin{subfigure}[b]{0.8\textwidth}
       \includegraphics[width=1\linewidth]{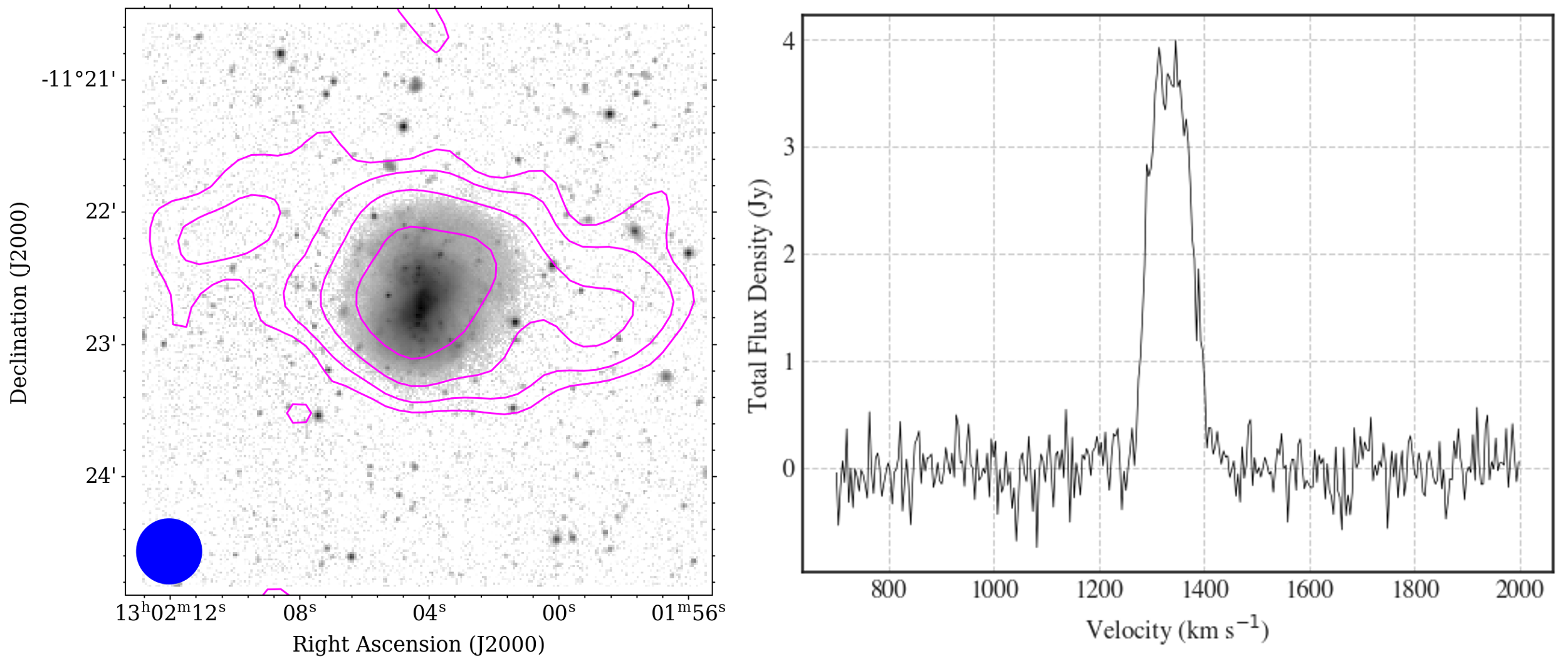}
   \caption{WALLABY J130204-112248 (NGC~4290).  The magenta \HI\ moment zero contour levels are at (1.8, 3.6, 7.2 and 14.4) $\times 10^{20}$ ~cm$^{-2}$.}
   \label{fig:223}
\end{subfigure}

\begin{subfigure}[b]{0.8\textwidth}
   \includegraphics[width=1\linewidth]{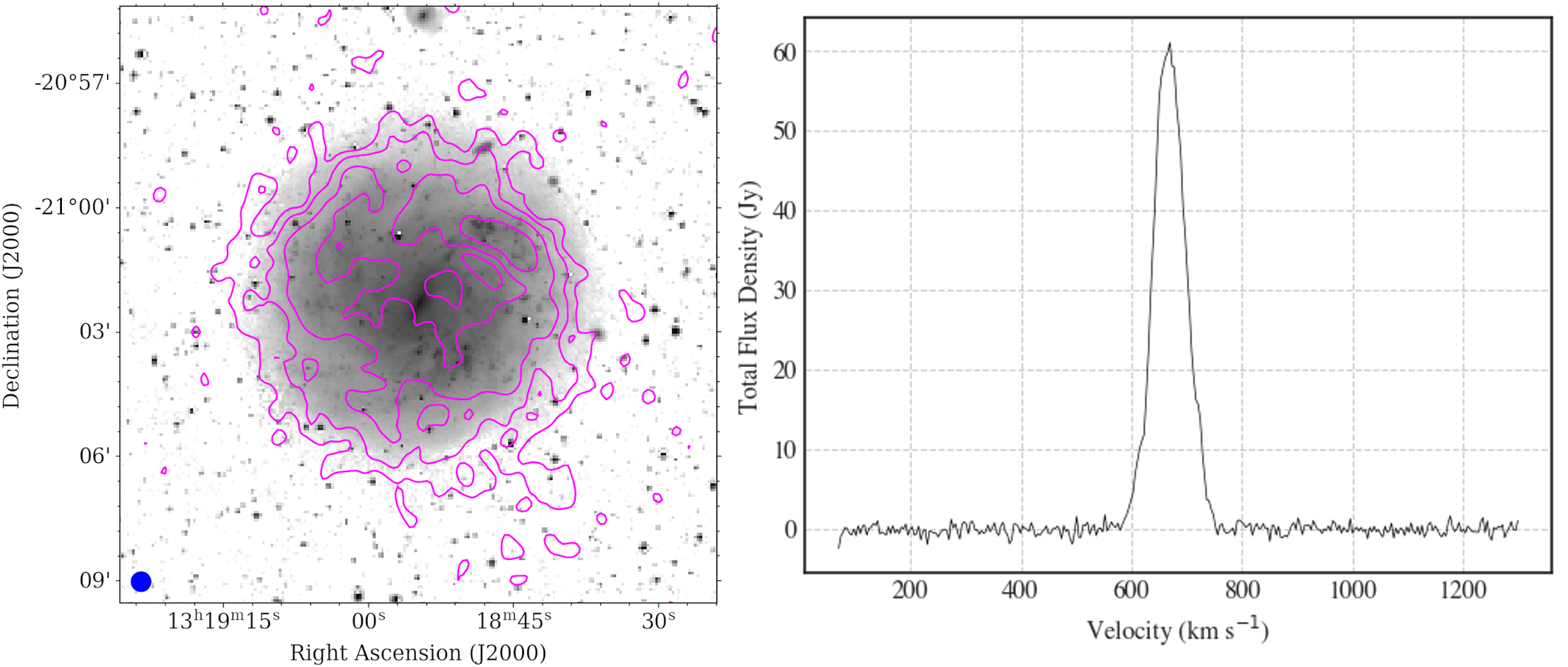}
   \caption{WALLABY J131851-210223 (NGC~5068).  The magenta \HI\ moment zero contour levels are at (2.5, 5.0, 10.0 and 20.0) $\times 10^{20}$ ~cm$^{-2}$.}
   \label{fig:210}
\end{subfigure}

\begin{subfigure}[b]{0.8\textwidth}
   \includegraphics[width=1\linewidth]{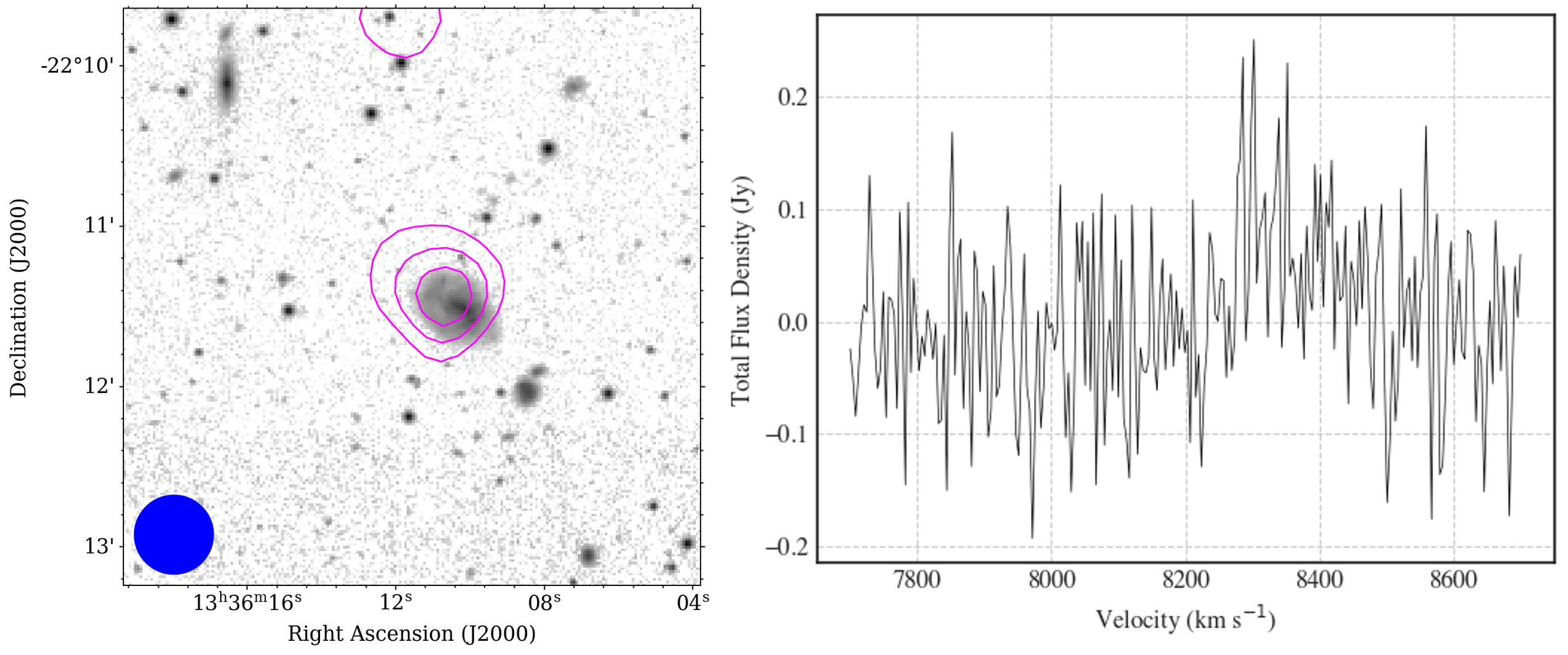}
   \caption{WALLABY J133610-221131. New \HI\ detection associated with LEDA~817885.  The magenta \HI\ moment zero contour levels are at (1.0, 2.0 and 3.0) $\times 10^{20}$ ~cm$^{-2}$.}
   \label{fig:50} 
\end{subfigure}

\caption{\HI\ sources identified by our model that are not catalogued in the default 30-arcsecond WALLABY DR2 catalogue. The left column shows the \HI\ moment zero column density maps as magenta contours overlaid on $g$-band images from the Legacy Survey. The higher-density regions are closer to the centre. The synthesised beam is shown in the bottom left corner of each moment zero map.  The right column shows the integrated \HI\ spectrum for each source. }\label{fig:newsources}
\end{figure*}

Figure~\ref{fig:newsources} presents the \HI\ moment zero column density maps across the entire emission line in the left column and the integrated spectra in the right column.  A summary of the observed properties of these three additional \HI\ sources within the NGC~5044 pointings can be found in Table~\ref{tab:newsource}.  The \HI\ spectral line parameters were measured using {\tt MBSPECT} function within the {\tt MIRIAD} software package \citep{Sault95}. Our results here highlight the value of our model and the invaluable role that such automated systems can play in improving efficiency, cross-checking and augmenting current source-finding workflows for very large surveys.

\section{Discussion}
\subsection{Implications of our results for large \HI\  surveys}
The machine-learning based workflow that we describe in this paper builds upon and leverages the strengths of SoFiA --- a source-finding tool that is well-understood and widely-used within the \HI\ community \citep[e.g.\ ][]{koribalski2020wallaby,hartley23}.  As described by \citet{serra2015SoFiA} and \citet{westmeier2021SoFiA}, SoFiA works well when the data are relatively clean and have Gaussian noise characteristics.  However, in the presence of non-Gaussian noise where the noise is a combination of imaging systematics and residual continuum subtraction or calibration artefacts, as shown in Fig. 6 in \citet{leahy2019askap}, it is more difficult for SoFiA to disentangle true detections from noise and artifacts. This is especially the case for low SNR sources, which will result in the cataloguing of a large number of false positives within the output candidate catalogue.

To this end, the combination of SoFiA and the 3D CNN-based model that we present here provides a source-finding method that is more capable at differentiating between true \HI\ sources from false positives due to non-Gaussian noise properties, relative to using SoFiA on its own. Admittedly human verification is still required in the current source-finding workflow, however, the addition of our model to SoFiA significantly reduces the number of false positive detections. The reduction in the number of false positive detections at low SNR leads to greater source-finding efficiency for the very large datasets that are generated by the next-generation \HI\ surveys, such as the WALLABY survey.  Our results also provide strong support for the use of multiple source-finding methods in order to optimise and maximise the output from very large surveys.  As we progress towards the SKA era of large surveys, results from SKA source-finding challenges based on simulated datasets may also not reveal the true challenge that is ahead when real observational data becomes available.
 
\subsection{Limitations of our method}
The machine-learning based method presented here is not an end-to-end source-finding tool and works in a complementary manner and leverages the strengths of SoFiA.  The advantage is that we are building on a well-understood source-finding tool and the contribution from our method is to enhance and further automate the functions of SoFiA. As such, our method is more interpretable and reproducible; and less of a `black box'. The model that we have presented here is a proof-of-concept and there are clear avenues for enhancement and expansion.

Using simulated \HI\ sources, we verify the potential efficacy and efficiency of the proposed machine-learning based method presented here.  However, a key result is that the model's accuracy on real datasets does not match its performance on simulated data.  As described in Section~4.2, our model is not able to recover the entire set of confirmed DR2 sources.  We show that SNR alone is insufficient to fully characterise the non-linear properties of both observational datasets and that of our model.  As such the range of properties spanned by true \HI\ detections needs to be better sampled within the training set of the model.  To include sources with rarer and a larger range of properties, a much larger dataset will be required than the ones used in this paper.  We also demonstrated that a more complex model with more convolutional layers does not translate to a significant improvement in performance. Hence, the size and diversity of the training dataset will ultimately drive future improvements to our method.

{The central focus of our future work will be to broaden the scope and diversity of the training dataset. This will involve not only the inclusion of much larger data samples to provide a richer learning experience for the model but also a deliberate emphasis on exploring objects which occupy a much larger range of observed properties such as lower SNR, broader line widths and larger angular extents. By integrating more examples of sources with rarer properties, the model's ability to accurately identify and classify objects in a wider range of conditions will be substantially improved.} We also note that as we aim to recover more FN sources and reduce the number of FP, we have to ensure the robustness of the model by preventing any possibility of over-fitting.   

{At the current stage, the accuracy of our model heavily relies on the labeling capabilities of human experts. In low SNR scenarios, our model theoretically can only approach, but not surpass, the accuracy of human experts. We are exploring methods to enable our model to exceed human expert accuracy even without better labels.}

Through these targeted efforts, we anticipate significant strides in our model's capability to analyze complex and large volumes of \HI\ datasets. Future improvements to our proposed method will make it a more robust resource for enabling accurate and comprehensive source-finding from very large surveys.

\section{Conclusion}
As SKA pathfinder surveys such as WALLABY get underway, there is a pressing need for increased automation in data analysis processes such as source finding from large data cubes.  Manual source-finding by astronomers is no longer a sustainable method given the data rates and volumes expected from surveys such as WALLABY.  To this end, we present a proof-of-concept machine-learning based workflow that works in a complementary manner to SoFiA.  Linear source-finding algorithms such as those used by SoFiA's smooth-and-clip do not perform well for data cubes which exhibit complex or non-Gaussian noise properties --- many false positive candidate detections are generated.

In this paper, we demonstrate that our workflow performs reasonably well using both a simulated and real WALLABY DR2 datasets.  Our model exhibits high accuracy in distinguishing between actual \HI\ sources and noise, even in challenging real-world conditions characterized by lower SNR and the presence of various processing artifacts. In summary, the key contributions of our work are as follows:
\begin{itemize}
    \item We developed a 3D Convolutional Neural Network model, specifically tailored to process three-dimensional \HI\ data. This model efficiently handles both two-dimensional spatial and one-dimensional spectral information inherent in data cubes, and leverages
    the correlated nature of true \HI\ detections in the spectral dimension. 
    \item Working alongside the SoFiA software, our model processes intermediate products (candidate list) generated by SoFiA and effectively reduces a substantial number of false positives.
    \item As an added bonus, we report the discovery of a new \HI\ source in LEDA~817885, further demonstrating the value of our approach.  More generally, such a discovery also argues for the use of multiple source-finding methods. 
    \item While focused on radio astronomical data, the methodology has potential applications in other areas of astronomy where multidimensional data is prevalent.
\end{itemize}

The quantitative analysis, supported by confusion matrices and experiments results, reveals the model's strengths and limitations.
Although the performance on real data does not completely match the near-perfect results obtained from the simulated dataset, our workflow still represents a significant advancement in the field of astronomical data analysis, considering the inherent complexities of real-world data.  This research paves the way for future studies and developments in source-finding from large surveys where manual analysis is impractical and unsustainable. By automating the initial stages of data filtering, our method allows astronomers to concentrate their efforts on the most promising data, thereby enhancing the efficiency and productivity of their research.

\printendnotes
\begin{acknowledgement}
This scientific work uses data obtained from Inyarrimanha Ilgari Bundara, the CSIRO Murchison Radio-astronomy Observatory. We acknowledge the Wajarri Yamaji People as the Traditional Owners and native title holders of the Observatory site. CSIRO’s ASKAP radio telescope is part of the Australia Telescope National Facility (https://ror.org/05qajvd42). Operation of ASKAP is funded by the Australian Government with support from the National Collaborative Research Infrastructure Strategy. ASKAP uses the resources of the Pawsey Supercomputing Research Centre. Establishment of ASKAP, Inyarrimanha Ilgari Bundara, the CSIRO Murchison Radio-astronomy Observatory and the Pawsey Supercomputing Research Centre are initiatives of the Australian Government, with support from the Government of Western Australia and the Science and Industry Endowment Fund. Parts of this research were conducted by the Australian Research Council Centre of Excellence for All Sky Astrophysics in 3 Dimensions (ASTRO 3D) through project number CE170100013
\end{acknowledgement}

\printbibliography

\end{document}